\documentclass[a4paper,12pt]{article}
\usepackage{lmodern}    %Latin Modern字体
\usepackage{graphicx}   %图片支持
\usepackage{amsmath}    %数学宏包
\usepackage{amsthm}     %排版定理
\usepackage{amssymb}    %数学符号，例：等号上加三角
\usepackage{abstract}
\usepackage{cite}      %参考文献多个引用合并
\usepackage{color}      %改变字体颜色 \textcolor{color}{text}
\usepackage{ulem}       %添加下划线 \uline{text}
\usepackage{enumerate}  %列表环境 \begin{enumerate}[间距选项]     \end{enumerate}
\usepackage{enumitem}   %列表环境中调节各种间距
\usepackage{tikz}       %画图
\usepackage{mathrsfs}
\usetikzlibrary{arrows.meta} %使用tikz库中的箭头Stealth
\usepackage{caption}    %添加图表标题
\usepackage[top=2.5cm,bottom=2.3cm,left=2.5cm,right=2.2cm]{geometry}%页边距
\usepackage[colorlinks,linkcolor=blue,anchorcolor=blue,citecolor=blue]{hyperref} %交叉引用 编号超链接
\numberwithin{equation}{section} %使公式编号带有章节号
\usepackage{booktabs}   %排版表格，\toprule、\midrule和\bottomrule命令用以排版三线表的三条线
\usepackage{longtable}  %提供排版跨页长表格的longtable环境，替代tabular命令
\usepackage{threeparttable} % 可以在表格之后增加表格注释
\allowdisplaybreaks[3]  %在align环境中允许公式跨页 括号中的参数可为1,2,3,4。数值越大,执行跨页的强度越大.
\usepackage{authblk}    %作者环境

\theoremstyle{definition}

\usepackage{xcolor}

\usepackage{ragged2e}
\justifying\let\raggedright\justifying % 两端对齐 (不加此命令行内公式无法自动换行)

\title{The symmetric (2+1)-dimensional Lotka--Volterra equation with  self-consistent sources}

\author[1]{Mengyuan Cui}
\author[1,\footnote{5572@cnu.edu.cn}]{Chunxia Li}
\author[2]{Yuqin Yao}
\affil[1]{School of Mathematical Sciences, Capital Normal University, Beijing 100048, China}
\affil[2]{College of Science, China Agricultural University, Beijing 100083, China}
\date{}

\begin{document}

\maketitle

 \begin{abstract}
The symmetric $(2+1)$-dimensional Lotka--Volterra equation with self-consistent sources is constructed and solved by employing the source generation procedure, whose solutions are expressed in terms of pfaffians. As special cases of the pfaffian solutions, different types of explicit solutions are obtained, including dromions, soliton solutions and breather solutions.
 \end{abstract}
% \begin{abstract}
% We have formulated non-commutative equations and obtained a primary solution and a secondary solution. Furthermore, by establishing a connection between pseudo-determinants and determinants, we have derived a set of three equations.
% \end{abstract}
% \noindent{\bf Keywords}: non-commutaitve Toda lattice with self-consistent
% sources, quasideterminant solutions, constant variation

% % REQUIRED
% \begin{MSCcodes}
% 37K60, 37K40, 68T07
% % 37K60 Lattice dynamics; integrable lattice equations
% % 37K40 Soliton theory, asymptotic behavior of solutions of infinite-dimensional Hamiltonian systems
% % 68T07 Artificial neural networks and deep learning
% % 37M15 Discretization methods and integrators (symplectic, variational, geometric, etc.) for dynamical systems
% % 39A14 Partial difference equations
% \end{MSCcodes}

\section{Introduction}
In recent years, there has been extensive research on (2+1)-dimensional integrable systems \cite{ablowitz1991solitons,konopelchenko1993solitons,konopelchenko2013introduction}. In particular, one crucial aspect is to explore new (2+1)-dimensional soliton equations. In various fields such as fluid dynamics, nonlinear optics, particle physics, general relativity, differential and algebraic geometry, and topology, several well-known examples of multi-dimensional integrable systems have been identified. Currently, there are a couple of effective methods for discovering (2+1)-dimensional integrable systems. One of these methods is to find integrable extensions of known (2+1)-dimensional integrable systems. For example, two coupled KP equations were discovered in two different research directions for the well-known KP equation. One is the so-called KP equation with self-consistent sources. and the other is generated through what is now called ``Pfaffianization" \cite{hirota1991hierarchies,hirota2004direct}. 
Following the leading work by Mel'nikov \cite{mel1983equations,mel1987direct,mel1988exact,mel1989capture,mel1989interaction,mel1990integration,mel1992integration}, much attention has been paid to soliton equations with self-consistent sources (SESCSs). A number of methods have been developed to study SESCSs, such as inverse scattering method, Darboux transformation, Hirota's bilinear method, dressing method and squared eigenfunction symmetry method \cite{deng2003multisoliton,hu2007new,lin2001solving,shao2005solutions,xiao2004generalized,yong2018lump,yun2002integral,zeng2000integration,zeng2001two,zeng2003negaton,zhang2003n,zhang2003n1,liu2009generalized}.
Lately, Hu and Wang suggested the source generation procedure which provides an efficient and unified way to construct and solve SESCSs \cite{hu2006construction,hu2007new,wang20072d,wang2007pfaffian,wang2008nizhnik}. The source generation procedure is in essence variation of constants and has been successfully applied to different types of soliton equations.

%Discrete integrable systems are of growing interest. 
In literature, some work has been done on discrete soliton equations with sources. In \cite{gegenhasi2006integrable}, integrability of the differential-difference KdV equation with a source was investigated. In \cite{liu2005toda}, the extended Toda lattice hierarchy was constructed by squared symmetric eigenfunctions, for which the non-autonomous Darboux transformation was derived. Furthermore, the two-dimensional Toda lattice equation, discrete KP equation and the semi-discrete BKP equation have been extended to their corresponding equations with self-consistent sources by source generation procedure, along with which determinant solutions and pfaffian solutions are derived, respectively \cite{hu2006construction,wang20072d,wang2007pfaffian}.
%Up to this point, the research on discrete soliton equations with sources and self-consistent sources has been gradually developing. 

It is well known that Lotka--Volterra (LV) equation 
\begin{align}\label{1+1LV}
    u_t(n)+e^{u(n)+u(n+1)}-e^{u(n)+u(n-1)}=0
\end{align}
is one of the most important lattices. In \cite{villarroel1996volterra,gilson2003two,dai2003decomposition,inoue1999construction}, several (2+1)-dimensional generalizations of equation \eqref{1+1LV} are presented. 
%Additionally, there exist two (2+1)-dimensional (two discrete and one continuous) extensions of the Lotka--Volterra lattice \cite{gilson2003two,inoue1999construction}. 
In \cite{hu2004integrable}, the symmetric (2+1)-dimensional Lotka–Volterra (2DLV) equation is proposed together with its bilinear B{\"a}cklund transformation, Lax pair and Pfaffian solutions. Moreover, explicit
solutions including dromions and soliton solutions are derived from the pfaffian solutions for the symmetric $2$DLV equation. As is explained in \cite{hu2004integrable}, the property of strong two-dimensionality seems to be closely related to the existence of dromions which has been proved to be true for the DS equation \cite{hu2008source,benney1969wave,davey1974k}, the NVN equation \cite{nizhnik1980integration,novikov1986two} and the symmetric $2$DLV equation. In this paper, we shall apply the source generation procedure to construct and solve the symmetric $2$DLV equation with self-consistent sources ($2$DLV ESCS). It will be very interesting to explore the corresponding explicit solutions such as dromions as well.  

%The most well-known example is the Davey–Stewartson (DS) equations [1, 2] which strongly generalize the nonlinear Schrodinger equation. Another interesting example is the Loewner–Konopelchenko–Rogers (LKR) equations [3]which strongly generalize the sine-Gordon equation. For the Korteweg–de Vries (KdV)equation, the most famous (2+1)-dimensional generalization, the Kadomtsev–Petviashvili(KP) equation [4], is only a weak generalizaton. Physically the KP equation arises insituations where one-dimensional motion governed by the KdV equation is weakly perturbed  in a perpendicular direction. A second, less well studied, generalization of the KdV equationis the Nizhnik–Novikov–Veselov (NNV) equation [5, 6]ut + uxxx + uyyy + 3(φxxu)x + 3(φyyu)y = 0, u = φxy . (1)On the line y = x, (1) reduces to the KdV equation. In this way one sees that (1) is a stronggeneralization of the KdV equation.The property of strong two-dimensionality seems to be closely related to the existenceof localized, exponentially decaying solutions (dromions). Indeed, this is the key featurewhich leads to the existence of the underlying plane-wave structure of such solutions in theDS equations [7–9] and in the NVN equation [10–12]. To our knowledge it seems thatexamples of strong (2+1)-dimensional generalizations of integrable equations shown in theliterature are restricted to the continuous case. Therefore, it is quite natural to consider strong(2+1)-dimensional generalizations of discrete integrable equations.

This paper is organized as follows. In Section \ref{section2}, by using the  source generation procedure, the symmetric $2$DLV ESCS as well as its DKP-type pfaffian solutions are presented.
In Section \ref{section3}, explicit solutions of the symmetric $2$DLV ESCS including
dromions, soliton solutions and breather solutions, are derived from the pfaffian solutions. Section \ref{conclusion} is devoted to conclusions and discussions.

\section{The symmetric \texorpdfstring{2}{}DLV ESCS and DKP-type pfaffian solutions} 

\label{section2}

The symmetric $2$DLV equation reads as \cite{hu2004integrable}
\begin{equation}\label{LV equ}
    \begin{aligned}
         2u_t&+e^{u+\Delta_m^2\phi_n}-e^{-u+\Delta_n^2\phi}+e^{u+\Delta_n^2\phi_m}
      -e^{-u+\Delta_m^2\phi}
       +e^{-u+\Delta_n^2\phi_{m\bar{n}}}-e^{u+\Delta_m^2\phi_{\bar{m}}}
       \\&\qquad+e^{-u+\Delta_m^2\phi_{\bar{m}n}}
       -e^{u+\Delta_n^2\phi_{\bar{n}}}=0,\quad u=\Delta_m\Delta_n\phi
    \end{aligned}
\end{equation}
 where $u=u(m,n,t)$, $\phi=\phi(m,n,t)$, the subscript $t$ denotes partial derivative as usual and the subscripts involving the discrete variables $m$ or $n$ denote
shifts:
%: $u_n=u(m,n + 1,t)$ and $u_{\bar n}=u(m, n-1,t)$For instance,
\begin{align*}
    u_m\equiv u(m+1,n,t),\quad u_{\bar{n}} \equiv u(m,n-1,t), \quad u_{\bar{m}n} \equiv u(m-1,n+1,t).
\end{align*}
The $\Delta_m$ and $\Delta_n$ are standard difference operators defined by
\begin{align*}
    \Delta_mu=u_m-u,\quad \Delta_nu=u_n-u.
\end{align*}
In the case that $m=n$, the symmetric $2$DLV equation \eqref{LV equ} reduces to \eqref{1+1LV}. In this sense, \eqref{LV equ} is regarded as a strong generalization of \eqref{1+1LV}.

Through the dependent variable transformation
\begin{align}\label{trans}
    u=\ln \frac{f_{mn}f}{f_mf_n},
\end{align}
equation \eqref{LV equ} is transformed into the multilinear form
\begin{align}\label{multilinear-LV}
   \begin{split}
   & \sinh\left(\frac{1}{2}D_n\right)\left[\left(D_te^{\frac{1}{2}D_m}-e^{D_n-\frac{1}{2}D_m}+e^{D_n+\frac{1}{2}D_m}\right)f\cdot f\right]\cdot\left(e^{\frac{1}{2}D_m} f \cdot f\right)\\
   &\qquad\qquad +\sinh\left(\frac{1}{2}D_m\right)\left[\left(D_te^{\frac{1}{2}D_n}-e^{D_m-\frac{1}{2}D_n}+e^{D_m+\frac{1}{2}D_n}\right)f\cdot f\right]\cdot\left(e^{\frac{1}{2}D_n} f \cdot f\right)=0,
    \end{split}
\end{align}
where the bilinear operators $D_t^k$ and exp($D_n$) are defined by \cite{
hirota2004direct}
\begin{equation*}
        D_t a \cdot b=(\partial_t-\partial_{t'})a(t)b(t')|_{t'=t},\quad
    \exp(\delta D_n)a(n)\cdot b(n)=a(n+\delta)b(n-\delta).
    % &\sinh{(\delta D_n)}=\dfrac{e^{\delta D_n}-e^{-\delta D_n}}{2}.
\end{equation*}
% By introducing $x$ and $y$ such that $\partial_x+\partial_y=2\partial_t$, 
% \eqref{multilinear-LV} is decoupled into the following bilinear form
% \begin{align*}
%     (D_xe^{\frac{1}{2}D_m}-e^{D_n-\frac{1}{2}D_m}+e^{D_n+\frac{1}{2}D_m})f\cdot f&=0\\
%     (D_ye^{\frac{1}{2}D_n}-e^{D_m-\frac{1}{2}D_n}+e^{D_m+\frac{1}{2}D_n})f\cdot f
%     &=0.
% \end{align*}
The multilibear equation \eqref{multilinear-LV} has the DKP-type Pfaffian solution %\cite{hu2004integrable}
\begin{equation*}
    f=(1,2,\cdots,2N)
\end{equation*}
where the Pfaffian elements $(i,j)$ are determined by the relations
\begin{align}
    &(i,j)_n=(i,j)+\theta_{i,n}\theta_j-\theta_i\theta_{j,n},\label{ijlv1}\\
    &(i,j)_m=(i,j)-\theta_{i,m}\theta_j+\theta_i\theta_{j,m},\label{ijlv2}\\
    &(i,j)_t=\dfrac{1}{2}(\theta_{i,\bar{n}}\theta_{j,n}-\theta_{i,n}\theta_{j,\bar{n}}+\theta_{i,m}\theta_{j,\bar{m}}-\theta_{i,\bar{m}}\theta_{j,m})\label{ijlv3}
\end{align}
and $\theta_i$ $(i=1,2,\cdots,2N)$ satisfy the linear dispersion relations
\begin{align}
&\theta_{i,mn}+\theta_i=\theta_{i,m}+\theta_{i,n},\label{theta1}\\
    &\theta_{i,t}=\dfrac{1}{2}(\theta_{i,\bar{n}}-\theta_{i,n}+\theta_{i,\bar{m}}-\theta_{i,m}).\label{theta2}
\end{align}
For simplicity, the index $d_i^j$ is introduced and defined by
\begin{align*}
(d_i^j,k)=\theta_k(m+i,n+j),\quad (d_i^j,d_k^l)=0,
\end{align*}
 so that the \eqref{ijlv1}-\eqref{ijlv3} can be written as 
\begin{align*}
    &(i,j)_{n}=(i,j)+(d_0^0,  d_0^1,i,j),\\
    &(i,j)_{m}=(i,j)+(d_1^0,d_0^0,i,j),\\
    &(i,j)_{t}=\frac{1}{2}((d_0^1,d_0^{-1},i,j)+(d_{-1}^0,d_1^0,i,j)).
\end{align*}

With the known pfaffian solution $f$ of the multilinear equation \eqref{LV equ}, we now construct the symmetric 2DLV ESCS. Following the source generation procedure \cite{hu2006construction}, we change $f$ into the following form 
\begin{align}\label{tau}
    \tau=(1,2,\cdots,2N),
\end{align}
whose Pfaffian elements are defined by
\begin{align}\label{pfaff}
    &(i,j)_{n}=(i,j)+(d_0^0,  d_0^1,i,j),\\
    &(i,j)_{m}=(i,j)-(d_1^0,d_0^0,i,j),\label{pfaff1}\\
    &(i,j)_{t}=\frac{1}{2}(\dot{C}_{i,j}(t)+(d_0^1,d_0^{-1},i,j)+(d_{-1}^0,d_1^0,i,j)),\label{pfaff2}
\end{align}
where $\dot{C}_{i,j}(t)$ is the $t$-derivative of $C_{i,j}(t)$ satisfying
\begin{align}\label{cij}
    C_{i,j}(t)=
\begin{cases}
    C_i(t),\quad\quad  i<j \quad and\quad j=2N+1-i,\quad 1\le i\le K\le N\\
    c_{i,j},\,\,\,\,\, \qquad  i<j \quad and \quad j\neq 2N+1-i. 
\end{cases}
\end{align}
%where $C_i(t)$ is an arbitrary function of $t$. 

It is obvious that $\tau$ no longer satisfies the symmetric $2$DLV equation \eqref{multilinear-LV} since $C_{i,j}(t)$ becomes dependent of $t$. In fact, we can prove that $\tau$ given by \eqref{tau} satisfies the new equation
\begin{equation}
    \begin{split}\label{scs1}
    &\sinh\left(\frac{1}{2}D_n\right)\left[\left(D_te^{\frac{1}{2}D_m}-e^{D_n-\frac{1}{2}D_m}+e^{D_n+\frac{1}{2}D_m}\right)\tau\cdot \tau-\sum_{i=1}^K\sinh\left(\frac{1}{2}D_m\right)h_i \cdot g_i\right]\\
&\qquad\qquad\cdot\left(e^{\frac{1}{2}D_m} \tau\cdot \tau\right)
    +\sinh\left(\frac{1}{2}D_m\right)\biggl[\left(D_te^{\frac{1}{2}D_n}-e^{D_m-\frac{1}{2}D_n}+e^{D_m+\frac{1}{2}D_n}\right)\tau\cdot \tau\\
    &\qquad\qquad-\sum_{i=1}^K\sinh\left(\frac{1}{2}D_n\right)g_i \cdot h_i\biggl]
    \cdot\left(e^{\frac{1}{2}D_n} \tau \cdot \tau\right)=0
    \end{split}
    \end{equation}
    with 
\begin{align}\label{gi}
    g_i&=\sqrt{\dot{C}_i(t)}(d_0^0,1,\cdots,\hat{i},\cdots,2N),\quad i=1,2,\cdots,K,\\
h_i&=\sqrt{\dot{C}_i(t)}\label{hi}(d_0^0,1,\cdots,
    \widehat{2N+1-i},\cdots,2N),\quad i=1,2,\cdots,K.
\end{align}
Meanwhile, $\tau$, $g_i$ and $h_i$ satisfy the following two equations
    \begin{align}
    &e^{\frac{1}{2}D_m+\frac{1}{2}D_n}g_i\cdot \tau=\left(e^{\frac{1}{2}D_m-\frac{1}{2}D_n}-e^{-\frac{1}{2}D_m-\frac{1}{2}D_n}+e^{-\frac{1}{2}D_m+\frac{1}{2}D_n}\right)g_i\cdot \tau,\label{scs2}\\
    &e^{\frac{1}{2}D_m+\frac{1}{2}D_n}h_i\cdot \tau=\left(e^{\frac{1}{2}D_m-\frac{1}{2}D_n}-e^{-\frac{1}{2}D_m-\frac{1}{2}D_n}+e^{-\frac{1}{2}D_m+\frac{1}{2}D_n}\right)h_i\cdot \tau\label{scs3}.
\end{align}

%In order to demonstrate this conclusion, we may rewrite equations \eqref{pfaff}-\eqref{pfaff2} as 
%\begin{align*}
  %  (i,j)_{n}&=(i,j)+(d_0^0,d_0^1,i,j),\\
   % (i,j)_{m}&=(i,j)+(d_1^0,d_0^0,i,j),\\
   % (i,j)_{t}&=\dfrac{1}{2}(\dot{C}_{i,j}(t)+(d_0^1,d_0^{-1},i,j)+(d_{-1}^0,d_1^0,i,j)),
%\end{align*}
Actually, by detailed calculations we have
\begin{align}
    &(i,j)_{\bar{n}}=(i,j)+(d_0^0,d_0^{-1},i,j),\qquad
    (i,j)_{\bar{m}}=(i,j)+(d_{-1}^0,d_0^0,i,j),\label{ij1}\\
     &(i,j)_{mn}=(i,j)+(d_1^0,d_0^1,i,j),\qquad
     (i,j)_{m\bar{n}}=(i,j)+(d_1^0,d_0^{-1},i,j),\label{ij2}\\
     &(i,j)_{\bar{m}n}=(i,j)+(d_{-1}^0,d_0^1,i,j)\label{ij3}.
\end{align}
These results are then used to give expressions for the derivatives and differences of
$\tau$, $g_i$ and $h_i$. For simplicity, denote $(1,2,\cdots,2N)=(\bullet)$, $(1,\cdots,\hat{i},\cdots,\widehat{2N+1-i},\cdots,2N)=(\circ)$, 
$(1,\cdots,\hat{i},\cdots,2N)=(\star)$ and $(1,\cdots,\widehat{2N+1-i},\cdots,2N)=(
\ast)$ for short and extend these notations to write $(d_0^1,d_1^0,1,2,\cdots,2N)= (d_0^1,d_1^0,\bullet)$ and so on. Using equations \eqref{ij1}-\eqref{ij3}, we obtain
\begin{align*}
    &\tau_n=(\bullet)+(d_0^0,d_0^1,\bullet),\quad\tau_m=(\bullet)+(d_1^0,d_0^0,\bullet),\quad\tau_{\bar{n}}=(\bullet)+(d_0^0,d_0^{-1},\bullet),\\
    &\tau_{\bar{m}}=(\bullet)+(d_{-1}^0,d_0^0,\bullet),\quad \tau_{mn}=(\bullet)+(d_1^0,d_0^1,\bullet),\quad \tau_{m\bar{n}}=(\bullet)+(d_1^0,d_0^{-1},\bullet),\\
    &\tau_{\bar{m}{n}}=(\bullet)+(d_{-1}^0,d_0^{1},\bullet), \quad\tau_t=\frac{1}{2}((d_0^1,d_0^{-1},\bullet)+(d_{-1}^0,d_1^0,\bullet))+\frac{1}{2}\sum_{i=1}^K\dot{C}_i(t)(\circ),\\
         &\tau_{m,t}=\frac{1}{2}((d_0^1,d_0^{-1},\bullet)+(d_0^{-1},d_0^{0},\bullet)-(d_0^{1},d_0^{0},\bullet)-(d_2^{0},d_0^{0},\bullet)+(d_1^{0},d_0^{-1},\bullet)\\
       &\qquad\,\,\,\,-(d_1^{0},d_0^{1},\bullet)+(d_0^1,d_0^{-1},d_1^0,d_0^{0},\bullet))
       +\frac{1}{2}\sum_{i=1}^K\dot{C}_i(t)((\circ)+(d_1^0,d_0^0,\circ)),      \\
         &\tau_{n,t}=\frac{1}{2}((d_{-1}^{0},d_1^0,\bullet)
         +(d_{-1}^{0},d_{0}^{1},\bullet)-(d_{-1}^{0},d_{0}^{1},\bullet)+(d_{0}^{0},d_{-1}^{0},\bullet)-(d_{0}^{0},d_{1}^{0},\bullet)\\&\qquad\,\,\,\,-(d_{0}^{0},d_{0}^{2},\bullet)+(d_{-1}^{0},d_{1}^{0},d_{0}^{0},d_{0}^{1},\bullet))
         +\frac{1}{2}\sum_{i=1}^K\dot{C}_i(t)((\circ)+(d_0^0,d_0^1,\circ)).\\
             &g_{i,n}=(d_0^1,\star),\quad g_{i,m}=(d_1^0,\star),\quad g_{i,mn}=(d_1^0,\star)+(d_0^1,\star)-(d_0^0,\star)
     +(d_0^1,d_1^0,d_0^0,\star),\\
     &h_{i,n}=(d_0^1,\ast),\quad h_{i,m}=(d_1^0,\ast),\quad h_{i,mn}=(d_1^0,\ast)+(d_0^1,\ast)-(d_0^0,\ast)
     +(d_0^1,d_1^0,d_0^0,\ast).
\end{align*}
On one hand, by direct substitution, %we only need to calculate $D_te^{\frac{1}{2}D_m}$ and $D_te^{\frac{1}{2}D_n}$ to obtain the following identities
equation \eqref{scs1} turns into the combination of the following two Pfaffian identities
\begin{align*}
   &(d_1^0,d_0^0,\circ)(\bullet)-(d_1^0,d_0^0,1,\bullet)(\circ)=(d_1^0,\ast)(d_0^0,\star)-(d_1^0,\star)
   (d_0^0,\ast),\\
   &(d_0^1,d_0^0,\circ)(\bullet)-(d_0^1,d_0^0,1,\bullet)(\circ)=(d_0^1,\ast)(d_0^0,\star)-(d_0^1,\star)
   (d_0^0,\ast).
\end{align*}
On the other hand, substituting these results into \eqref{scs2} and \eqref{scs3} will lead to 
\begin{equation*}
\begin{aligned}
(d_0^1,d_1^0,d_0^0,\star)(\bullet)=&(d_0^1,\star)(d_1^0,d_0^0,\bullet)-(d_1^0,\star)(d_0^1,d_0^0,\bullet)+(d_0^0,\star)(d_0^1,d_1^0,\bullet),\\
(d_0^1,d_1^0,d_0^0,\ast)(\bullet)=&(d_0^1,\ast)(d_1^0,d_0^0,\bullet)-(d_1^0,\ast)(d_0^1,d_0^0,\bullet)+(d_0^0,\ast)(d_0^1,d_1^0,\bullet),
\end{aligned}
 \end{equation*}
%and
%\begin{equation*}(d_0^1,d_1^0,d_0^0,\ast)(\bullet)=(d_0^1,\ast)(d_1^0,d_0^0,\bullet)-(d_1^0,\ast)(d_0^1,d_0^0,\bullet)+(d_0^0,\ast)(d_0^1,d_1^0,\bullet),
%\end{equation*}
%respectively, 
respectively, which are nothing but Pfaffian identities.
%we have 
%\begin{align*}
    %&g_{i,n}=(d_0^1,\star),\quad g_{i,m}=(d_1^0,\star),\\
     %&g_{i,mn}=(d_1^0,\star)+(d_0^1,\star)-(d_0^0,\star)
     %+(d_0^1,d_1^0,d_0^0,\star),\\
     %&h_{i,n}=(d_0^1,\ast),\quad h_{i,m}=(d_1^0,\ast),\\
    %&h_{i,mn}=(d_1^0,\ast)+(d_0^1,\ast)-(d_0^0,\ast)
   %  +(d_0^1,d_1^0,d_0^0,\ast).
%\end{align*}
%Denote 
%$(d_i^j,1,\cdots,\hat{i},\cdots,2N)=(d_i^j,\star)$ and $(d_i^j,1,\cdots,\widehat{2N+1-i},\cdots,2N)=(d_i^j,
%\ast)$.

To sum up, equations \eqref{scs1}, \eqref{scs2} and \eqref{scs3} constitute a coupled system with
$K$ pairs of self-consistent sources which can be viewed as the symmetric $2$DLV ESCS. At the same time, $\tau$, $g_i$ and $h_i$ given by \eqref{tau}, \eqref{gi} and \eqref{hi} provide the associated Pfaffian solutions. 
%These results suggest that equations \eqref{scs1}, \eqref{scs2} and \eqref{scs3} hold true. Therefore, the functions $\tau$, $g_i$ and $h_i$ defined by \eqref{tau}, \eqref{gi} and \eqref{hi} can be considered as a DKP-type Pfaffian solution of equations \eqref{scs1}-\eqref{scs3}, which in turn represents the 2DLV ESCS in bilinear form.

% Next, we will decouple the multilinear equation \eqref{scs1} into bilinear form by introducing two auxiliary variables $x$ and $y$.
% \begin{align*}
%     (D_xe^{\frac{1}{2}D_m}-e^{D_n-\frac{1}{2}D_m}+e^{D_n+\frac{1}{2}D_m})f\cdot f&=\sum_{i=1}^K\frac{1}{2}(e^{\frac{1}{2}D_m}+e^{-\frac{1}{2}D_m})g_i \cdot h_i,\\
%     (D_ye^{\frac{1}{2}D_n}-e^{D_m-\frac{1}{2}D_n}+e^{D_m+\frac{1}{2}D_n})f\cdot f
%     &=\sum_{i=1}^K\frac{1}{2}(e^{\frac{1}{2}D_n}
%     +e^{-\frac{1}{2}D_n})g_i \cdot h_i,
% \end{align*}
% where $D_x+D_y=2D_t$.

With the help of the dependent variable transformations
\begin{align*}
    u=\ln{\frac{\tau_{mn}\tau}{\tau_m\tau_n}},\quad
    q_i=\frac{g_i}{\tau},\quad
    r_i=\frac{h_i}{\tau},
\end{align*}
equations \eqref{scs1}-\eqref{scs3} are transformed into the nonlinear symmetric 2DLV ESCS
\begin{align}\label{nonlinearLV}
  %  \begin{aligned}
    &2u_t+e^{u+\Delta_m^2\phi_n}-e^{-u+\Delta_n^2\phi}+e^{u+\Delta_n^2\phi_m}-e^{-u+\Delta_m^2\phi}
       +e^{-u+\Delta_n^2\phi_{m\bar{n}}}-e^{u+\Delta_m^2\phi_{\bar{m}}}
      +e^{-u+\Delta_m^2\phi_{\bar{m}{n}}}\nonumber\\
        &\qquad\qquad-e^{u+\Delta_n^2\phi_{\bar{n}}}=\frac{1}{4}\sum_{i=1}^K[(q_{i,mn}-q_i)(r_{i,m}-r_{i,n})+(q_{i,m}-q_{i,n})(r_i-r_{i,mn})],\\
    &q_{i,mn}e^{u}-q_{i,m}+q_ie^{u}-q_{i,n}=0,\qquad i=1,2,\cdots,K,\label{nc1}\\
    &r_{i,mn}e^{u}-r_{i,m}+r_ie^{u}-r_{i,n}=0,\qquad i=1,2,\cdots,K.\label{nc2}
%\end{aligned}
\end{align}
\section{Explicit solutions of the symmetric \texorpdfstring{2}{}DLV ESCS}\label{section3}
In the previous section, we have constructed the symmetric $2$DLV ESCS and obtained its Pfaffian solutions. In this section, we shall follow the the method in \cite{ohta1992pfaffian,athorne1991moutard} to derive $(M,N-M)$-dromions, $N$ soliton solutions and therefore multi-breather solutions for the symmetric $2$DLV ESCS.

Notice that \eqref{theta1} may be rewritten as
\begin{align*}
    \Delta_m\Delta_n\theta_i=0,
\end{align*}
which implies that each $\theta_i$ can be decomposed as
\begin{align}\label{decompose}
\theta_i(m,n,t)=\phi_i(n,t)+\psi_i(m,t).
\end{align}
Substituting \eqref{decompose} into \eqref{theta2}, we have
\begin{equation}    \label{phipsi}
\begin{aligned}
    2\phi_{i,t}&=\phi_{i,\bar{n}}-\phi_{i,n},\\
     2\psi_{i,t}&=\psi_{i,\bar{m}}-\psi_{i,m}.
\end{aligned}
\end{equation}
Based on the above calculations, the Pfaffian element $(i,j)$ in $f$ determined by \eqref{pfaff}, \eqref{pfaff1} and \eqref{pfaff2} can be established as
\begin{align}\label{ij33}
    (i,j)=C_{i,j}(t)+\phi_i\psi_j-\psi_i\phi_j+\int({\phi_{i,\bar{n}}\phi_{j,n}-\phi_{i,n}\phi_{j,\bar{n}}+\psi_{i,m}\psi_{j,\bar{m}}-\psi_{i,\bar{m}}\psi_{j,m}})dt.
\end{align}
% The simplest choices of these solutions are
% \begin{align*}
%     \phi_{i}=\alpha_i\left(\frac{1-p_i}{1+p_i}\right)^{-n}exp\left(\frac{-2p_it}{1-p_i^2}\right),\\
%     \psi_{i}=\beta_i\left(\frac{1-q_i}{1+q_i}\right)^{-m}\exp\left(\frac{-2q_it}{1-q_i^2}\right),
% \end{align*}
% in which $\alpha_i$, $\beta_i$, $p_i$ and $q_i$ are constants.

%From the equation \eqref{phipsi}, the $\phi_i$ and $\psi_i$ can be calculated. To find the solutions to the 2DLV ESCS, we  
Following the method proposed in \cite{ohta1992pfaffian}, we choose the following appropriate functions for $\phi_i$ and $\psi_i$ for the Pfaffian element $(i,j)$ in $f=(1,2,\cdots,2N)$. We take 
% \begin{align*}
%     \phi_i&=\alpha_i\left(\frac{1-p_i}{1+p_i}\right)^{-n}exp\left(\frac{-2p_it}{1-p_i^2}\right), &1\le i \le 2M, \\
%     \phi_i&=0,    &2M+1\le i\le 2N,\\
%     \psi_i&=0,   & 1\le i \le 2M, \\
%     \psi_i&=\beta_i\left(\frac{1-q_i}{1+q_i}\right)^{-m}\exp\left(\frac{-2q_it}{1-q_i^2}\right)\biggl|_{i=2N+1-i}, &2M+1\le i\le 2N.
% \end{align*}
 \begin{align*}
     \phi_i&=P_ie^{\eta_i}, &1\le i \le 2M, \\
     \phi_i&=0,    &2M+1\le i\le 2N,\\
    \psi_i&=0,   &1\le i \le 2M, \\
     \psi_i&=Q_{2N+1-i}e^{\xi_{2N+1-i}}, &2M+1\le i\le 2N
\end{align*}
where 
$$\eta_i=\dfrac{-2p_it}{1-p_i^2},\quad \xi_i=\dfrac{-2q_it}{1-q_i^2},\quad 
P_i=\alpha_i\left(\frac{1-p_i}{1+p_i}\right)^{-n},\quad 
Q_i=\beta_i\left(\frac{1-q_{i}}{1+q_{i}}\right)^{-m}$$
and $\alpha_i$, $\beta_i$, $p_i$ and $q_i$ are constants. With these assumptions, explicit solutions such as dromion solutions, soliton solutions and breather solutions are able to be derived.
%Based on the special cases, a series of solutions can be derived, such as dromion solutions, soliton sulutions and breather solutions. 
\subsection{Dromion solutions}
By taking $0<M<N$, we have from \eqref{ij33} that 
\begin{align*}
    &(i,j)=C_{i,j}(t)+\dfrac{p_i-p_j}{p_i+p_j}P_iP_je^{\eta_i+\eta_j},\,\,\,\quad\qquad\qquad\qquad\qquad \qquad\qquad\qquad1\le i<j\le 2M,\\
    &(i,2N+1-j)=C_{i,2N+1-j}(t)+P_iQ_{j}e^{\eta_i+\xi_{j}},\qquad\qquad\qquad\, 1\le i\le 2M,1\le j\le 2N-2M,\\
    &(2N+1-j,2N+1-i)=C_{2N+1-j,2N+1-i}(t)+\frac{q_i-q_j}{q_i+q_j}Q_iQ_je^{\xi_{i}+\xi_{j}},\,\, 1\le i<j\le 2N-2M.  
\end{align*}
In this case, $\tau$, $g_i$ and $h_i$ give the $M\times (N-M)$-dromion solutions.

\begin{figure}
    \centering
    \includegraphics[width=0.8\linewidth]{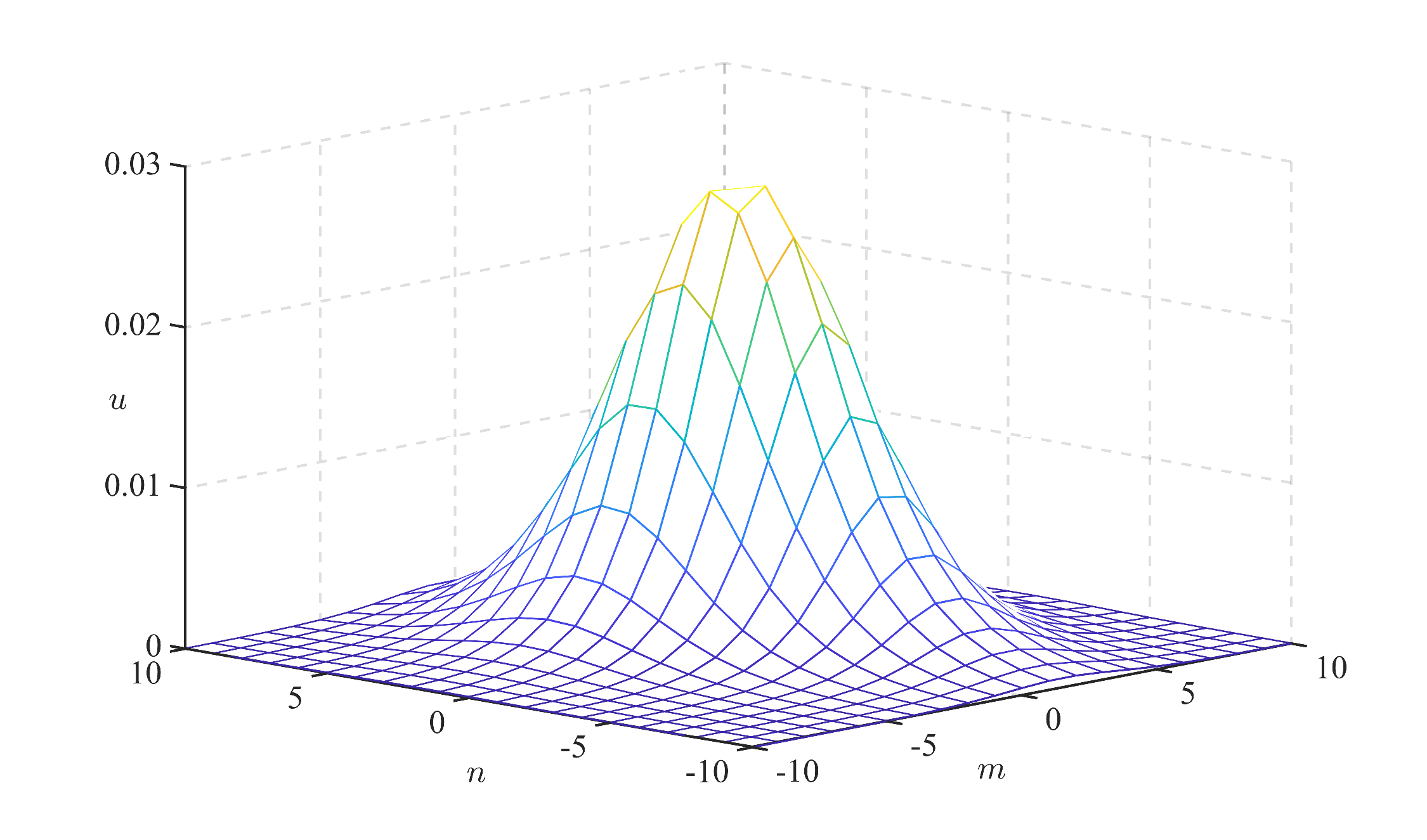}
    \caption{$(1,1)$-dromion solution of \eqref{nonlinearLV} with $t=1$, %\eqref{11dromion}, 
$C_1(t)=t^2,~C_2(t)=t,~\alpha_1=\alpha_2=\beta_1=\beta_2=1,~p_1=-\frac{1}{4},~q_2=\frac{1}{3},~c_{1,2}=-1,\\c_{1,3}=1,~c_{2,4}=-2,~c_{2,4}=-2$.}
    \label{fig:1}
\end{figure}

 % It is analogous to the propagation in the $m$ and $n$ directions of a two-soliton solution $f'$, $g$ and $h$. By setting $q_1=p_2=0$, we obtain
%     \begin{align*}
%         f''=\phi_1+\psi_2+\phi_1\psi_2+\mathcal{C}_{1,2}(t)\dfrac{1}{\psi_2},\qquad
%          g_1=\sqrt{2\dot{C}_1(t)}\psi_2,\qquad
%         h_1=\sqrt{2\dot{C}_1(t)}\phi_1.
%     \end{align*}
Consider the simplest case $M=1$, $N=2$ and $K=1$. According to the definition of $C_{i,j}(t)$ in \eqref{cij}, we have
% by setting $C_{1,3}(t)=C_{2,4}(t)=C_{1,4}(t)=C_{2,3}(t)=0$, we have
\begin{equation}\label{11dromion}
    \begin{aligned}
    \tau=&c_{1,2}c_{3,4}-c_{1,3}c_{2,4}+C_1(t)C_2(t)-c_{1,3}P_2Q_1e^{\eta_2+\xi_1}-c_{2,4}P_1Q_2e^{\eta_1+\xi_2}
\\&+C_2(t)P_1Q_1e^{\eta_1+\xi_1}+C_1(t)P_2Q_2e^{\eta_2+\xi_2}+c_{3,4}\dfrac{p_1-p_2}{p_1+p_2}P_1P_2e^{\eta_1+\eta_2}
    \\& +c_{1,2}\dfrac{q_1-q_2}{q_2+q_1}Q_1Q_2e^{\xi_1+\xi_2}
    +\dfrac{p_1-p_2}{p_1+p_2}\dfrac{q_1-q_2}{q_1+q_2}P_1P_2Q_1Q_2e^{\eta_1+\eta_2+\xi_1+\xi_2},\\
    g_1=&\sqrt{\dot{C}_1(t)}(C_2(t)Q_1e^{\xi_1}-c_{2,4}Q_2e^{\xi_2}+(c_{3,4}+\dfrac{q_1-q_2}{q_1+q_2}Q_1Q_2e^{\xi_1+\xi_2})P_2e^{\eta_2}),\\
    h_1=&\sqrt{\dot{C}_1(t)}(C_2(t)P_1e^{\eta_1}-c_{1,3}P_2e^{\eta_2}+(c_{1,2}+\dfrac{p_1-p_2}{p_1+p_2}P_1P_2e^{\eta_1+\eta_2})Q_2e^{\xi_2}).
\end{aligned}
\end{equation}
Furthermore, by setting $p_2=q_1=0$, we have
\begin{equation*}
    \begin{aligned}
    \tau=&c_{1,2}c_{3,4}-c_{1,3}c_{2,4}-c_{1,3}\alpha_2\beta_1+C_1(t)C_2(t)+(c_{3,4}\alpha_2+\beta_1C_2(t))P_1e^{\eta_1}\\&
    +(\alpha_2C_1(t)-c_{1,2}\beta_1)Q_2e^{\xi_2}-(\alpha_2\beta_1+c_{2,4})P_1Q_2e^{\eta_1+\xi_2},\\
    g_1=&\sqrt{\dot{C}_1(t)}(\alpha_2c_{3,4}+\beta_1C_2(t)-(\alpha_2\beta_1+c_{2,4})Q_2e^{\xi_2}),\\
     h_1=&\sqrt{\dot{C}_1(t)}(-c_{1,3}\alpha_2+c_{1,2}Q_2e^{\xi_2}+C_2(t)P_1e^{\eta_1}+\alpha_2P_1Q_2e^{\eta_1+\xi_2}),
\end{aligned}     
\end{equation*}
 which gives the $(1,1)$-dromion solution of the symmetric $2$DLV ESCS \eqref{nonlinearLV} (see Fig.\ref{fig:1}). 
 % The result may be readily extended to the general $M\times(N-M)$-dromion solution.
 \subsection{Soliton solutions}
By taking $2M=N$ and $C_{i,j}(t)=\delta_{i,2N+1-j}C_{i}(t)$, we have
\begin{align*}
    &(i,j)=\dfrac{p_i-p_j}{p_i+p_j}P_iP_je^{\eta_i+\eta_j},&1\le i<j\le N,\\
    &(i,2N+1-j)=\delta_{i,j
    }C_{i}(t)+P_iQ_je^{\eta_i+\xi_j},&1\le i, j\le N,\\
    &(2N+1-j,2N+1-i)=\frac{q_i-q_j}{q_i+q_j}Q_iQ_je^{\xi_i+\xi_j},&1\le i<j\le N.  
\end{align*}
In this case, $\tau$, $g_i$ and $h_i$ give the $N$-soliton solution.

\begin{figure}
    \centering
    \includegraphics[width=0.8\linewidth]{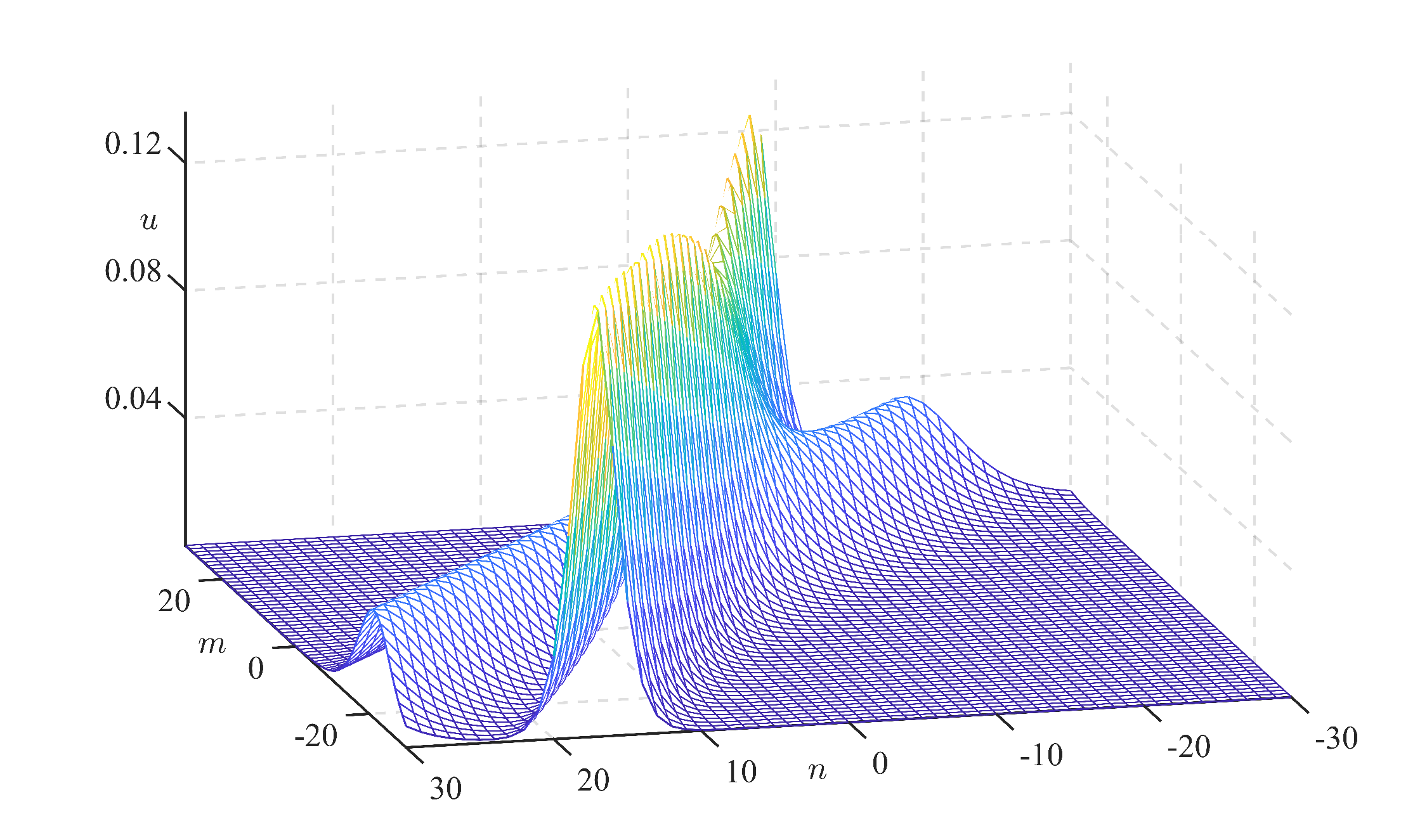}
    \caption{$2$-soliton solution of \eqref{nonlinearLV} with $t=2, ~C_1(t)=t^2,~C_2(t)=t,~\alpha_1=\alpha_2=\beta_1=\beta_2=1,~p_1=\frac{1}{5},~p_2=\frac{1}{2},~q_1=\frac{1}{6},~q_2=\frac{1}{4}$.}
    \label{fig:2}
\end{figure}

 Consider the case $N=2$ and $K=1$, we can obtain the 2-soliton solution (see Fig. \ref{fig:2})
\begin{equation}\label{2soliton}
    \begin{aligned}
    &\tau=C_{1}(t)C_{2}(t)+C_{1}(t)P_2Q_2e^{\eta_2+\xi_2}+C_{2}(t)P_1Q_1e^{\eta_1+\xi_1}
    \\&\qquad+\dfrac{p_1-p_2}{p_1+p_2}\dfrac{q_1-q_2}{q_1+q_2}P_1P_2Q_1Q_2e^{\eta_1+\eta_2+\xi_1+\xi_2},\\
    &g_1=\sqrt{\dot{C}_1(t)}(C_{2}(t)Q_1e^{\xi_1}+\dfrac{q_1-q_2}{q_1+q_2}P_2Q_1Q_2e^{\eta_2+\xi_1+\xi_2}),\\
    &h_1=\sqrt{\dot{C}_1(t)}(C_{2}(t)P_1e^{\eta_1}+\dfrac{p_1-p_2}{p_1+p_2}P_1P_2Q_2e^{\eta_1+\eta_2+\xi_2}).
\end{aligned} 
\end{equation}
Note that if we further set $p_2=q_2=0$, we have
\begin{align*}
    &\tau=\alpha_2\beta_2C_{1}(t)+C_{1}(t)C_{2}(t)+(\alpha_2\beta_2+C_{2}(t))P_1Q_1e^{\eta_1+\xi_1},\\
    &g_1=\sqrt{\dot{C}_1(t)}(C_{2}(t)+\alpha_2\beta_2)Q_1e^{\xi_1},\\
    &h_1=\sqrt{\dot{C}_1(t)}(C_{2}(t)+\alpha_2\beta_2)P_1e^{\eta_1},
\end{align*}
which is the one-soliton solution.
\subsection{Breather solutions}
In the preceding subsections, we have  obtained soliton solutions for the symmetric 2DLV ESCS. In what follows, we are going to derive breather solutions from soliton solutions.

\begin{figure}
    \centering
    \includegraphics[width=0.8\linewidth]{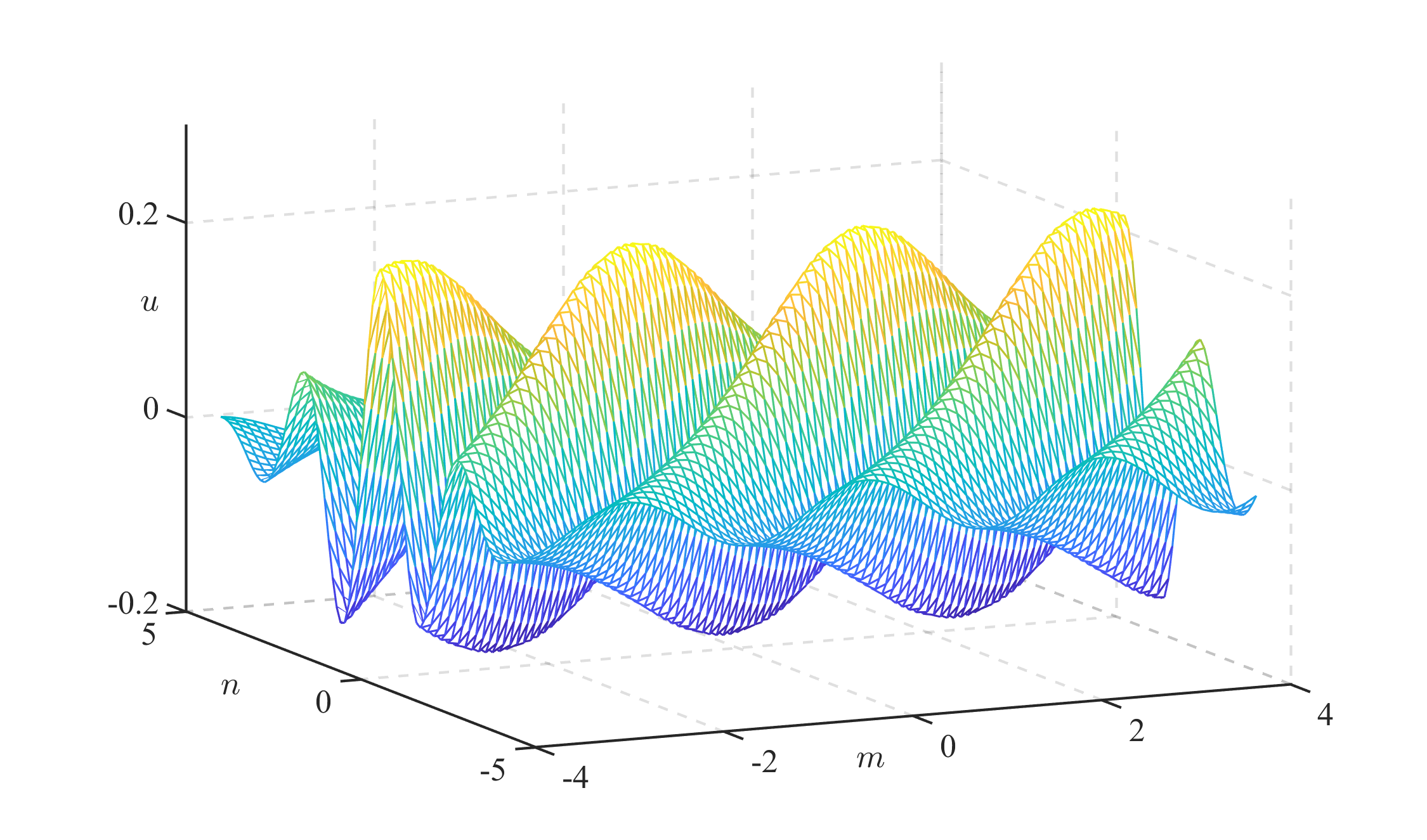}
    \caption{$1$-breather solution of \eqref{nonlinearLV} with $t=0.01, ~a=-1,~b=1,~c=-0.2,~d=-3,~\delta(t)=t^2+2,~\gamma(t)=-t^2-1$.}
    \label{fig:3}
\end{figure}

For the sake of convenience, we set $\alpha_i=\beta_i=1$. Consider the case $N=2$ and $K=1$. Let $^*$ denote complex conjugate. By taking $p_1=p_2^*=a+bi$, $q_1=q_2^*=c+di$ and $C_1(t)=C_2^*(t)=\gamma(t)+\delta(t)i$, we have the $1$-breather 
\begin{equation}\label{breather}
    \begin{aligned}
&\tau=\gamma^2(t)+\delta^2(t)+2\gamma(t)R(a,b)^{-\frac{n}{2}}R(c,d)^{-\frac{m}{2}}e^{2(I(a,b)+I(c,d))t}\cos(-\mathrm{Arg}(S(a,b))n
    \\&\qquad-\mathrm{Arg}(S(c,d))m-2(T(a,b)+T(c,d))t+\mathrm{Arg}(\delta(t)i))
    \\&\qquad+a_{12}R(a,b)^{-n}R(c,d)^{-m}e^{4(I(a,b)+I(c,d))t},\\
    &g_1=\sqrt{(\dot{\gamma}(t)+\dot{\delta}(t)i)}M(c,d)^{-m}e^{(I(c,d)-T(c,d)i)t}(\gamma(t)-\delta(t)i+\dfrac{di}{c}S(a,b)^{-n}\\
    &\qquad\,\,\, S(c,d)^{-m}e^{(I(a,b)+I(c,d)-(T(a,b)+T(c,d))i)t}),\\
    &h_1=\sqrt{(\dot{\gamma}(t)+\dot{\delta}(t)i)}M(a,b)^{-m}e^{(I(a,b)-T(a,b)i)t}(\gamma(t)-\delta(t)i+\dfrac{bi}{a}S(a,b)^{-n}\\
    &\qquad\,\,\, S(c,d)^{-m}e^{(I(a,b)+I(c,d)-(T(a,b)+T(c,d))i)t})
\end{aligned}
\end{equation}
where 
\begin{align*}
    R(x,y)=&\frac{(x-1)^2+y^2}{(x+1)^2+y^2},\quad  I(x,y)=\frac{2x}{x^2+y^2-1},\quad
    S(x,y)=\dfrac{-x^2-y^2+1+2yi}{(x+1)^2+y^2},\\ T(x,y)=&\frac{2y}{x^2+y^2-1},\quad
    M(x,y)=\dfrac{-x^2-y^2+1-2yi}{(x+1)^2+y^2},\quad a_{12}=-\frac{bd}{ac}
   %\quad a_{12}=-\frac{bd}{ac}&>1, \quad  \gamma \le 1,\quad  \gamma^2(t)+\delta^2(t)\ge 1,
\end{align*}
with $a,b,c,d$ being arbitrary real-valued constants (see Fig.\ref{fig:3}). Remarkably, $a_{12}>1$, $\gamma \le 1$ and $ \gamma^2(t)+\delta^2(t)\ge 1$
imply that $\tau$ is always positive, which accounts for the nonsingular solution. The general breather solutions can be obtained from the soliton solutions in a similar way.
\section{Conclusions and discussions}\label{conclusion}
The symmetric $2$DLV ESCS is constructed by employing the source generation procedure. DKP-type Pfaffian solutions of the system have also been
derived. In the case that $C_i(t)$ is independent of $t$, the pairs of sources $g_i$ and $h_i$ become zero identically. Consequently, the symmetric bilinear (or nonlinear) $2$DLV ESCS is reduced to the symmetric bilinear (or nonlinear) $2$DLV equation. Meanwhile, DKP-type Pfaffian solutions of the symmetric $2$DLV ESCS is reduced to the ones of the symmetric $2$DLV equation. Due to the property of strong two-dimensionality, the symmetric $2$DLV ESCS has been proved to have dromion solutions in addition to soliton solutions and breather solutions.

It has been shown that the NVN equation is a continuous analogue of the symmetric $2$DLV equation \cite{hu2004integrable}. Since the NVN equation has nonsingular lump solutions \cite{hu1996some,gegenhasi2020nonsingular}, it is believed that the symmetric $2$DLV equation and the symmetric $2$DLV ESCS should have nonsingular lump solutions too. We will discuss such problems elsewhere.

% \begin{exam}[dromion  solution]
%     The $(1,1)$-drimion solution is analogous to the propagation in the $m$ and $n$ directions of a two-soliton solution $f'$, $g$ and $h$. By setting $q_1=p_2=0$, we obtain
%     \begin{align*}
%         f''=\phi_1+\psi_2+\phi_1\psi_2+\mathcal{C}_{1,2}(t)\dfrac{1}{\psi_2},\qquad
%          g_1=\sqrt{2\dot{C}_1(t)}\psi_2,\qquad
%         h_1=\sqrt{2\dot{C}_1(t)}\phi_1.
%     \end{align*}
% \end{exam}
% This is the $(1,1)$-dromion solution of equations \eqref{scs1}-\eqref{scs3}. The general dromion of order  $(M,N)$, where $(M,N)$ is an even number, can be derived by considering the pfaffian $(1,2,...,M+N)$ with specific values assigned to the variables. In particular, we set $q_1=q_2=\cdots=p_{M+1}=\cdots=p_{M+N}=0$. For the case when M+N is odd, additional soliton solutions can be obtained by setting one more $p_i$ or $q_i$ equal to zero.

\section*{Acknowledgement}
This work was supported by the National Natural Science
Foundation of China (Grants Nos. 11971322 and 12171475).

%\appendix

\normalem %normal emphasis 防止因使用ulem包导致参考文献中期刊名出现下划线

\end{document}